\documentclass[12pt]{article}
\usepackage{graphicx}
\def\th{\theta}         
\def\ga{\gamma}         

       \def\Lam{\Lambda}
\def\al{\alpha}         \def\ep{\epsilon}
\def\de{\delta}         
\def\om{\omega}         \def\Om{\Omega}
        
\def\vphi{\varphi}

\title{The Formation of non-Keplerian Rings of Matter about Compact
  Stars} 

\author{S. P. Drake\thanks{Now at: Surveillance Systems Division,
    D.S.T.O., P.O. Box 1500, Salisbury S.A. 5108,
    Australia.}\\
  Department of Physics, University of Adelaide, Australia \\
  email: sam.drake@dsto.defence.gov.au}

\date{\today}

\begin{document}
\maketitle
\begin{abstract}
  The formation of energetic rings of matter in a Kerr spacetime with
  an outward pointing acceleration field does not appear to have
  previously been noted as a relativistic effect. In this paper we show
  that such rings are a gravimagneto effect with no Newtonian analog,
  and that they do not occur in the static limit. The energy
  efficiency of these rings can, depending of the strength of the
  acceleration field, be much greater than that of Keplerian disks.
  The rings rotate in a direction opposite to that of compact star
  about which they form. The size and energy efficiency of the
  rings depend on the fundamental parameters of the spacetime as
  well as the strength the acceleration field. 
\end{abstract}


\section{Introduction} \label{sec:i}  

A compact celestial body may accrete matter from a nearby companion.
If the accreting matter has large enough angular momentum, a potential
barrier will form stopping the in-fall.  Matter bouncing back from the
angular momentum barrier, will collide with the in-falling one and
eventually an equilibrium condition is reached where by most of matter
moves on circular orbits and is  confined to a plane.  Once
equilibrium is established, if matter is subjected to no forces other
than gravity, it moves in geodesic orbits.  Matter confined to a plane
and moving in geodesic circular motion is said to form a Keplerian
disk.  Keplerian disks adequately describe galactic motion since stars
are sufficiently far apart to be considered  non-interacting point
sources. On the other hand, matter surrounding a hot compact source of
radiation, is likely to be affected not only by the radiation pressure
of the source but also by its own internal forces. For these reasons
the accreted matter will not necessarily follow Keplerian orbits.

Despite this, not much work has been done on modeling non-Keplerian
disks, see~\cite{Bur98, Kin98}
for some examples.  In this paper we analyze the dynamics 
underlining the formation of rotating structures in the presence of an
acceleration field within the full theory of general relativity.
Although we limit our considerations to a point particle approach, the
fundamental results should also be manifest also in a fluidodynamical
treatment.

The first problem one faces in examining motion exterior to a rotating
mass is the lack of an exact solution to Einstein's equations. To
describe the dynamics in general relativity we must know the space-time
geometry.  Birkhoff's theorem~\cite{Wei72}
tells us that the
space-time exterior to a non-rotating, spherically symmetric,
electrically neutral configuration, is the Schwarzschild solution.
Unfortunately though, there is no generalization of  Birkhoff's
theorem for rotating stars. The space-time of a stationary, uncharged
and rotating black hole is uniquely described by Kerr solution
\cite{Cha83}.
This has led a number of authors~\cite{Kra78, McM91, Mag93, Mag95,
  Dra97}. 
to suggest that the exterior of a rotating star may be described to
sufficient accuracy by Kerr geometry.  Thus far however, nobody has
been able to match a ``physically sensible'' interior solution to the
Kerr metric.  Although one may think that Kerr metric still describes
the basic properties of a space-time exterior to a rotating star,
mainly stemming from stationarity and axisymmetry.

In this paper, we investigate the astrophysical importance of a
general relativistic effect arising in Kerr geometry which has no
Newtonian or Schwarzschild analogue. In Kerr space-time, one finds
that non-geodesic (spatially) circular orbits may have, at each value
of their coordinate radius, an extreme acceleration for non-zero  orbital
angular velocities (with respect to infinity). As we
shall see, this effect is responsible for the existence of narrow and
stable rings of matter, populated by highly energetic particles. In
Newtonian theory and Schwarzschild geometry acceleration extrema occur
only for zero angular velocity.

The existence of an extremal acceleration implies that for a range of
angular velocities, an increase in the {\it modulus} of the angular
velocity, requires a larger outward pointing acceleration to maintain
a circular orbit circular see.  This is contrary to the Newtonian case
were an increase in the modulus of angular velocity, requires a
smaller outward acceleration to maintain a circular orbit, in all
regions.

This effect was first noticed in the Schwarzschild space-time
by~\cite{Abr74}
and then in the Kerr metric by~\cite{Fel91}.
In the latter case the effect exists at all values of the coordinate
distance from a rotating source so one can even hope to measure it in
a weak field regime~\cite{Fel95}.

In Section~\ref{sec:cmitkm}, we shall outline the main properties of
accelerated circular orbits in the Kerr metric. In
Section~\ref{sec:nkr}, we show how ring structures form due to the
existence of an acceleration field and how energy considerations allow
us to decide about the stability of the rings. In Section~\ref{sec:ai}
we discuss their possible astrophysical importance.   Finally
in the last Section we summarize the results, draw our conclusions,
and discuss possible further work. 

In what follows we shall use geometrized units such that 
 $G = c= 1$, $G$ being the gravitational 
constant and  $ c $ the vacuum speed of light; Greek indices run from 0 to 3 
and signature is 
chosen as $+2$.

\section{Circular motion in the Kerr metric} \label{sec:cmitkm}

In Boyer-Lindquist coordinates, 
$x^{\al} = \{ t, r, \th, \phi \}$,  the Kerr metric is described by the line 
element
\begin{eqnarray}
  \label{eq:eqn1}
ds^2 & = &    
-\left(1 - {2Mr\over r^2 + a^2 \cos^2\th}\right) dt^2 - {4 a M r
  \sin^2\th\over r^2 + a^2 \cos^2\th}
dt d\phi \nonumber \\ 
& + & {(r^2 + a^2)^2 - a^2 \sin^2\th (r^2 + a^2 - 2Mr)\over r^2 + a^2
  \cos^2\th}\sin^2\th d\phi^2  \nonumber \\
& + & {r^2 + a^2
  \cos^2\th\over r^2 + a^2 - 2Mr}dr^2 + (r^2 + a^2 \cos^2\th) d\th^2.
  \end{eqnarray}
The constants  $M$ and $a$ are the mass and specific angular momentum 
of the black hole in units of length.

Matter confined in the equatorial plane and moving in spatially circular 
orbits 
has a four velocity
\begin{equation}
  \label{eq:eqn2}
u^\rho= {dx^\rho\over d\tau}= e^{\vphi} \left( \de^\rho_t + \Om\de^\rho_\phi 
\right),
\end{equation}
where $\tau$ is the proper-time along the orbits,  $\Om$ is 
the angular frequency of the orbital revolution as it would
be measured at  infinity, it has the dimensions of length$^{-1}$, 
$\delta^\rho_t$ and $\delta^\rho_\phi$ are Kronecker deltas. 
The quantity $e^{\vphi}$, known as the red-shift factor, is derived from the 
normality condition 
$u_\al u^\al = -1$, and reads:
\begin{equation}
  \label{eq:eqn3}
e^{\vphi}=\left[1-{2Mr\over r^2 + a^2 \cos^2\th}(1-a\Omega)^2-(r^2+a^2)
\Omega^2\right]^{-1/2}\, ,\quad \theta=\pi/2.  
\end{equation}
The four-acceleration of non-geodesic orbits is given by
\begin{equation}
  \label{eq:eqn4}
\dot u_\rho= u_{\rho;\sigma}u^\sigma, 
\end{equation}
where ``$\,;\,$'' denotes the covariant derivative relative with
respect to the metric. The dot above $u_\rho$ is used to denote the
absolute derivative with respect to the proper-time.  In the case of
circular orbits (\ref{eq:eqn2}) in a Kerr space-time (\ref{eq:eqn1})
the four acceleration is~\cite{Fel91, Fel94, Fel95}
\begin{equation}
  \label{eq:eqn5}
\dot u_\rho = {x\over M} {(\omega-\omega_{g+})(\omega - \omega_{g-})\over(\omega - \omega_{c+})
(\omega - \omega_{c-})}\de^r_\rho, \qquad \theta=\pi/2,
\end{equation}
where,
\begin{equation}
  \label{eq:eqn6}
\begin{array}{lll}
x& \equiv {M\over r}\,, \quad  \al \equiv {a\over M}\,, \quad
\omega \equiv {M\Om\over 1 - \al M \Om} \cr  
\Lam &= 1 + \al^2 x^2 - 2 x \,,\quad
\omega_{g\pm} = \pm x^{3/2}\,,  \quad  
\omega_{c\pm} = \al x^2 \pm x \sqrt{\Lam}\, . 
\end{array}
\end{equation}
The notation used here differs from that of cited references
in that inverse distances are measured and all the quantities are
dimensionless and scaled in terms of mass.  In what follows we shall
refer to $x$ as to a {\it position} or a {\it distance}, although
it is proportional to the inverse of the coordinate $r$.

Here $\omega$ is the scaled angular frequency of revolution,
$\omega=\omega_{g_\pm}$ are the geodesic orbits, and $\omega =
\omega_{c_\pm}$ are the causal boundary conditions, i.e, trajectories
with $\om > \om_{c+}$ or $\om < \om_{c-}$ have velocities faster than
light.  Using equations~(\ref{eq:eqn1}),~(\ref{eq:eqn4})
and~(\ref{eq:eqn5}) we define the scalar acceleration as 
\begin{equation}
  \label{eq:eqn7}
a_{cc}  \equiv (\dot u^\rho \dot u_\rho)^{1/2}=
{x\Lambda^{1/2}\over M}\left| {(\omega-\omega_{g+})(\omega - \omega_{g-})
\over(\omega - \omega_{c+})(\omega - \omega_{c-})}\right|,
\end{equation}
where $a_{cc}$ has the dimensions of a length$^{-1}$.  In what
follows, a positive acceleration refers to outward pointing.

\begin{figure}[htbp]
  \begin{center}
    \includegraphics[width=\textwidth]{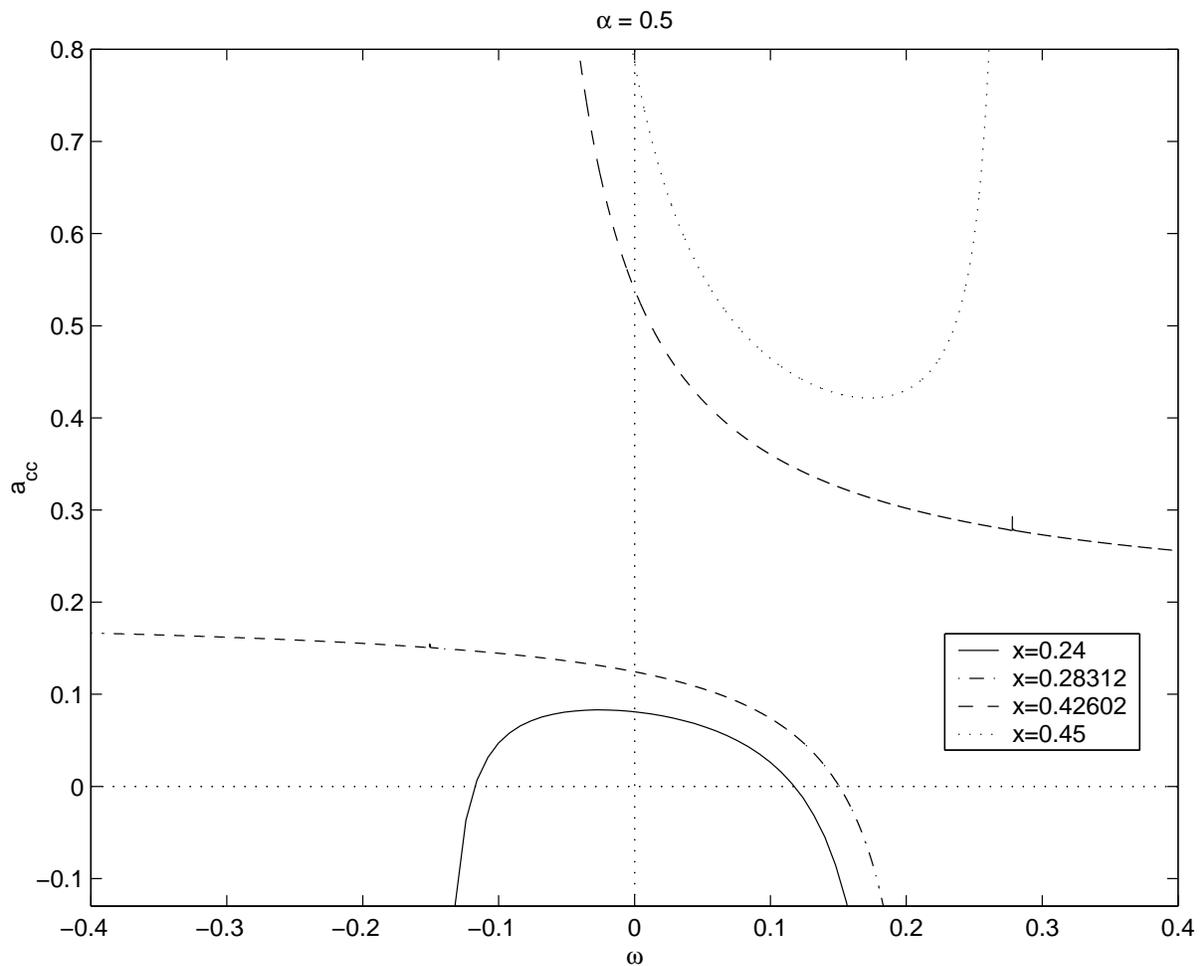}
    \caption[ale05.eps]{Acceleration required to keep orbits circular as a function of the scaled
      angular frequency. Plots are shown for different distances
      from the source, $x \equiv M/r = 0.24, 0.28312, 0.42602, 0.45$.  The
      Kerr parameter is $\al = 0.5$.\label{fig:ale05}}
      \end{center}
\end{figure}

\begin{figure}[htbp]
  \begin{center}
    \includegraphics[width = \textwidth]{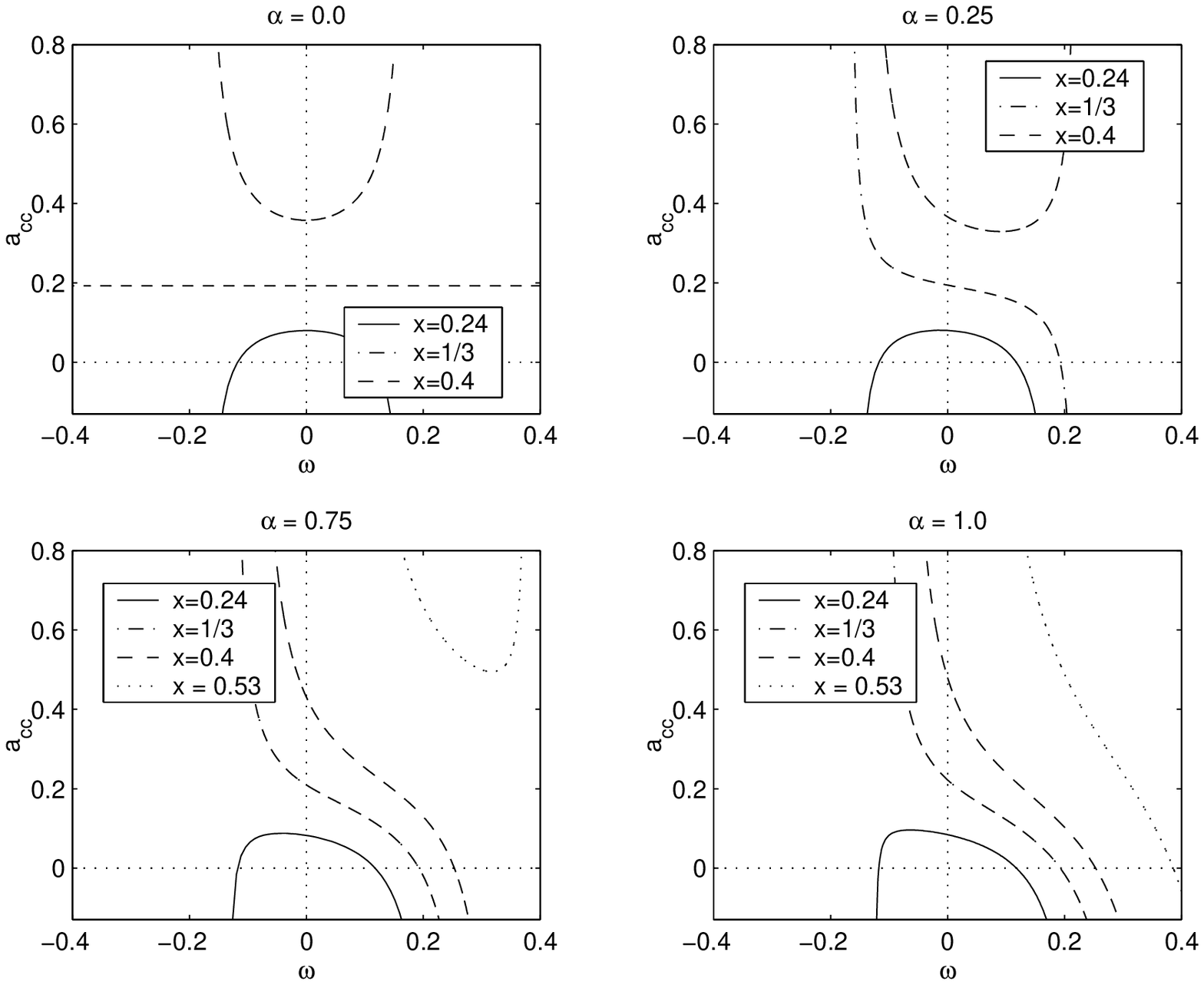}
    \caption[varal.eps]{Acceleration required to keep orbits circular as a function of the scaled
      angular frequency. Plots are shown for different distances
      from the source, $x \equiv M/r = 0.24, 0.33, 0.4, 0.53$.  The
      Kerr parameter is $\al = 0, 0.25, 0.75, 1.0$. \label{fig:varal}}
      \end{center}
\end{figure}

Equation~(\ref{eq:eqn7}) determines the  acceleration  required to
keep a particle or  fluid element, with an angular frequency
$\omega$, at a distance $x$. Figures \ref{fig:ale05}~and~\ref{fig:varal}
show plots of $a_{cc}$ as function of $\omega$ for different values of
$x$ and the rotation parameter $\alpha$.  In all of these graphs at
large distances from the stars centre (small values of $x$) the
acceleration has a maximum for small angular velocities ($\omega
\approx 0$). With the exception of non-rotating stars ($\al =0$), as
one gets closer to the centre, i.e.  as $x$ gets larger, it becomes
clear that the maximum acceleration occurs for negative values of
$\om$. A negative $\om$ denotes a rotation opposite to that of the
star. Moving closer still, we see that eventually acceleration has no
maximum or minimum. The furthest distance for which there is no
maximum is called $x_I$, we shall derive its value shortly. Closer to
the centre we reach a distance where the acceleration has a minimum,
the furthest distance at which this occurs is called $x_{II}$. 
\begin{figure}[htbp]
  \begin{center}
    \includegraphics[width = 7cm]{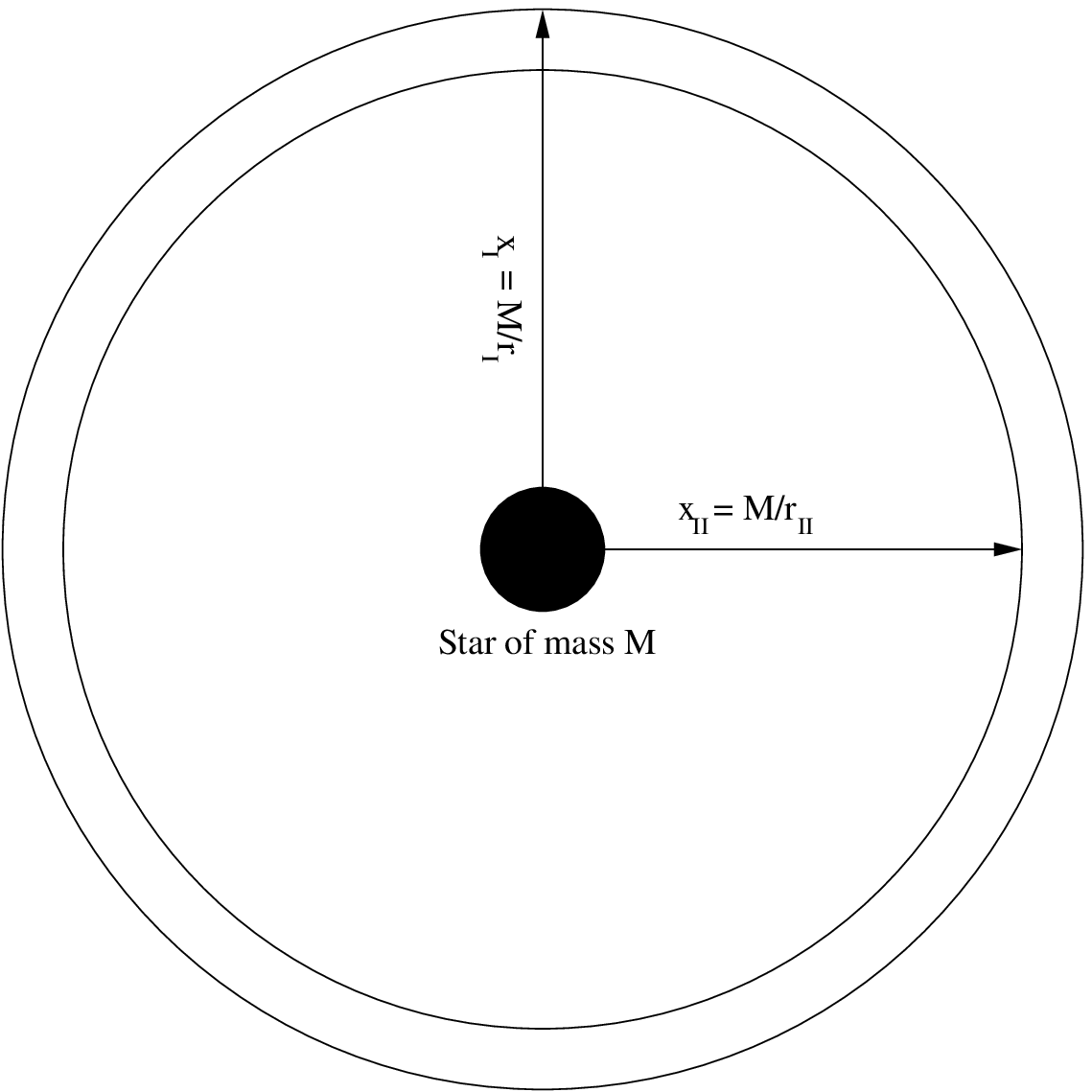}
     \end{center}
\end{figure}

This description motivates us to divide the equatorial plane into three
regions:
\begin{itemize}
\item Region 1; $0\le x< x_I$, acceleration has a maximum.
\item Region 2; $x_I \le x \le x_{II}$, acceleration has no extremal value.
\item Region 3; $x>x_{II}$
\end{itemize}

In figures~\ref{fig:ale05} and~\ref{fig:varal} $\|\al \| \le 1$ the
reason for this is that if $\al > 1$ the space-time has a naked
singularity, i.e., a singularity without an event horizon. The cosmic
censorship conjecture states that this does not occur in
nature~\cite{Wal84,Ger79,Pen79}.

\begin{figure}[htbp]
  \begin{center}
    \includegraphics[width=\textwidth]{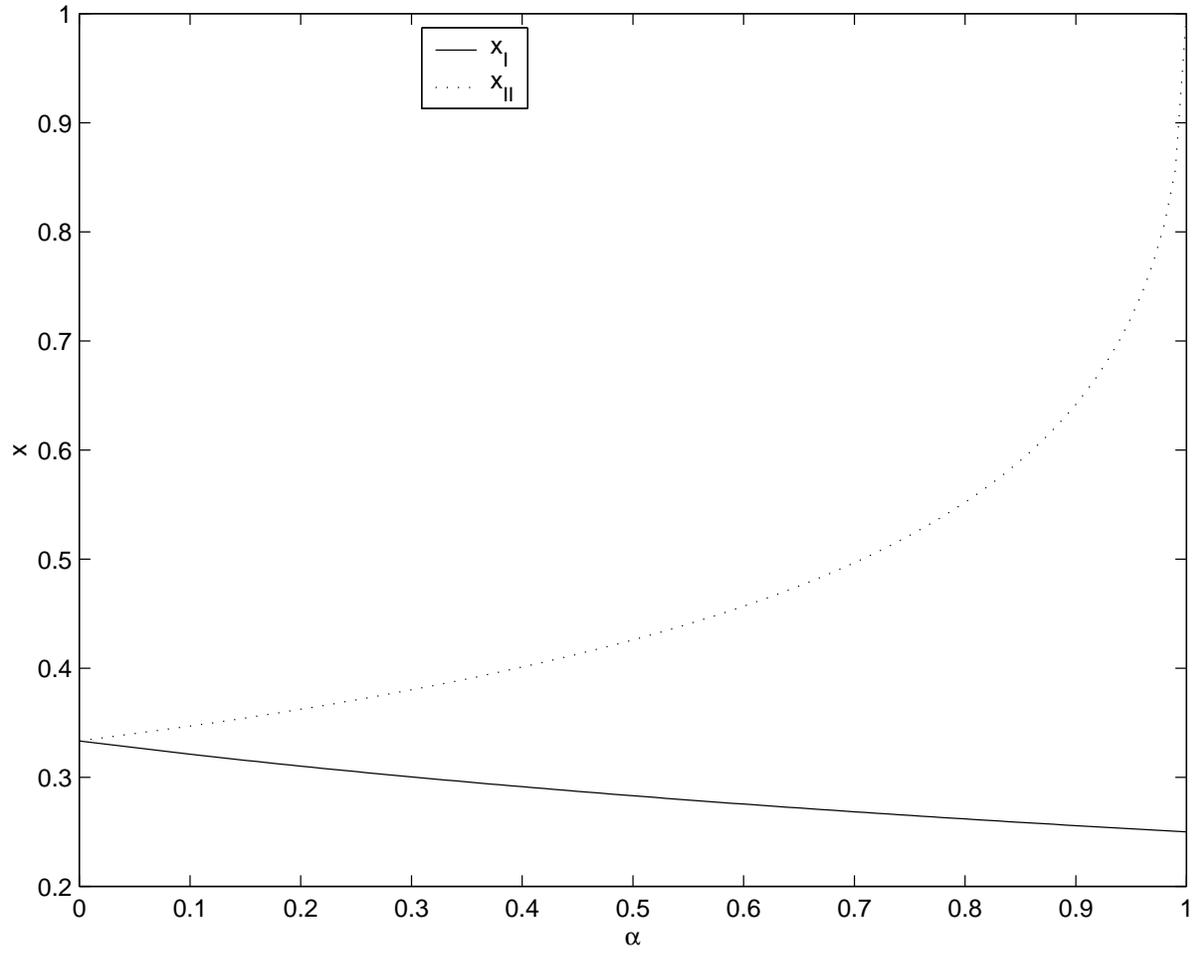}
    \caption[zerosof10.eps]{Region boundaries as a function of $\al$
    \label{fig:zerosof10}}
  \end{center}
\end{figure}

Until now we have focused on determining the acceleration required to
keep orbits circular for a range of angular velocities. Suppose we invert the
problem and determine angular velocity for known  accelerations. To do
this we rearrange equation~(\ref{eq:eqn7}),
\begin{equation}
  \label{eq:eqn8}
\omega_\pm(x;a_{acc})={x\over \chi-1}\left[-\alpha x\pm \sqrt{\alpha^2x^2
+(\chi-1)(-1+2x+\chi x)}\right]
\end{equation}
where $ \chi\equiv x\Lambda^{1/2}/a_{cc}$.
\begin{figure}[htbp]
  \begin{center}
    \includegraphics[width = \textwidth]{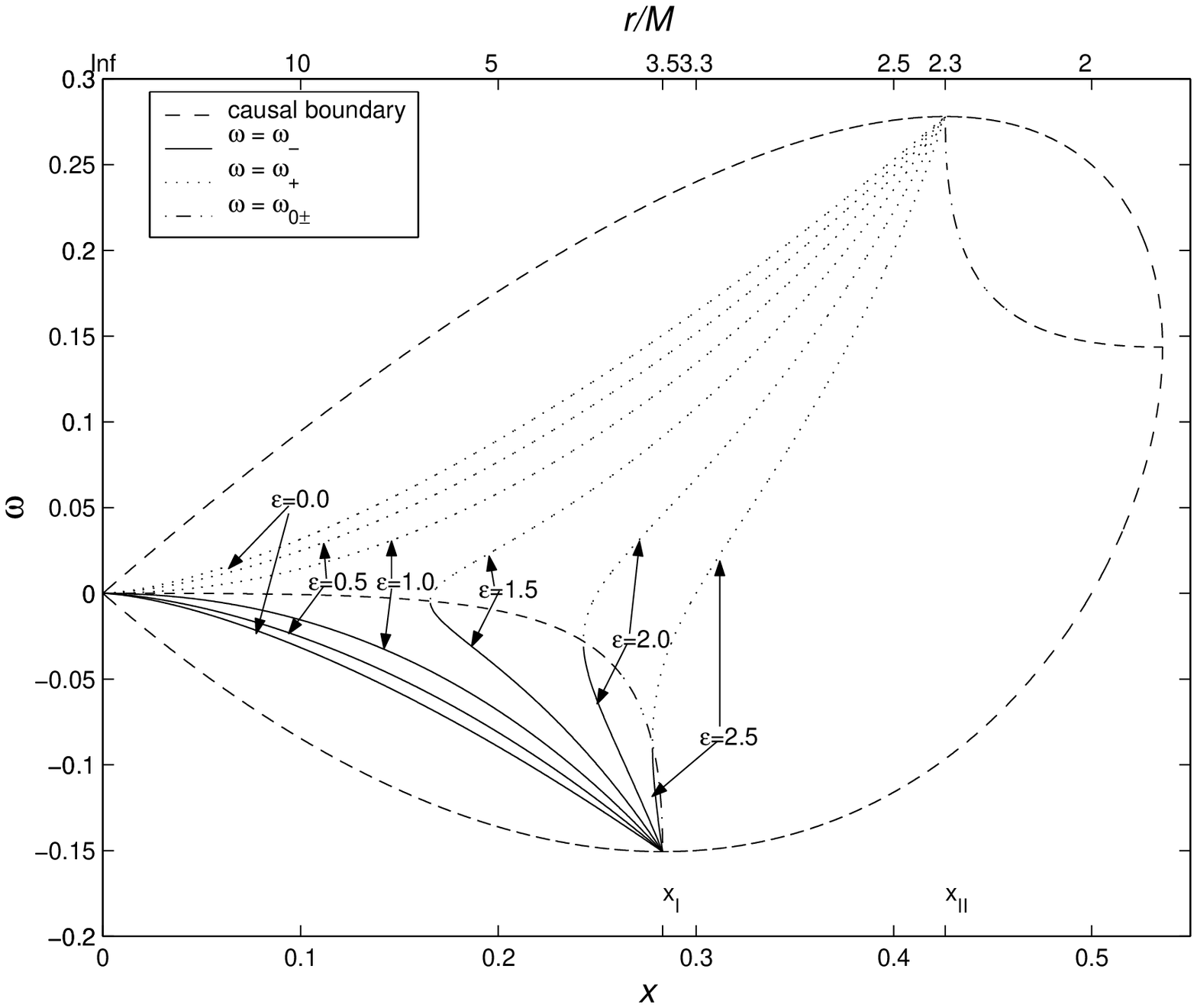}
    \caption[wvx.eps]{Angular velocity versus distance  for $a_{cc} = \ep x^2\sqrt{\Lam}/M$ at  $\ep =
      0.5, 1, 1.5, 2.0,2.5$ with $\al = 0.5$.\label{fig:wvx}}
  \end{center}
\end{figure}
Figure~\ref{fig:wvx} shows the angular velocity  of
circular orbits as a function of $x$ for $a_{cc} = \ep
x^2\sqrt{\Lam}$ at various values of $\ep$. The reasons for this choice of
acceleration field will become clear later. The permitted angular velocities
($\omega_{c-}<\omega<\omega_{c+}$), correspond to time-like orbits.
The solid branch of each curve corresponds to the $\omega_-$ solution
of~(\ref{eq:eqn8}), the dotted branch to $\omega_+$.  The family of
extremal accelerated circular orbits are shown by the curves
$\omega=\omega_{0_\pm}$ (to be introduced shortly). Those with
$\omega=\omega_{0_+}<0$ are maximally accelerated while those with
$\omega=\omega_{0_-}>0$ are minimally accelerated.  The maximum
(minimum) acceleration occur for $x<x_I$ ($x>x_{II}$), at $\om =
\om_{0+}$ ($\om=\om_{0-}$) where
\begin{equation}
  \label{eq:eqn10}
\om_{0\pm} = -{1\over 2\alpha}
\left[1 - 3x \mp \sqrt{(1-3x)^2 - 4\al^2 x^3}\right].  
\end{equation}
Which correspond to an acceleration 
\begin{equation}
  \label{eq:eqn11}
a_{0_\pm}\equiv  a_{cc}(\om_{0_\pm})  
 = {x\Lambda^{1/2}\over M}{\sqrt{(1 -3x)^2 - 4 \al^2 x^3}
\mp (1-3x)\over \sqrt{(1 -3x)^2 - 4 \al^2 x^3}\mp (1-3x + 2\al^2x^2)}.
\end{equation}
With the exception of the static limit ($\al = 0$), $\om_{0\pm} \neq
0$ for all $x>0$.  The real zeros of the argument of the square-root,
specify $x_I$ and $x_{II}$.  Figure~\ref{fig:zerosof10} is a plot of $x_I$
and $x_{II}$ versus $\al$. It can be shown analytically, or seen from
figure~\ref{fig:zerosof10} that in the Schwarzschild limit ($\al =0$)
$x_I$ and $x_{II}$ occur at the same distance, $x_I = x_{II} = 1/3$.
As $\al$ decreases $x_I$ ($r_I$ increases) and $x_{II}$ increases
($r_{II}$ decreases) until $\al = 1$ at which point $x_I = 1/4$ and
$x_{II} = 1$.

\section{Non-Keplerian rings}\label{sec:nkr}

In this section we show that, under certain conditions, stable rings
of matter form about compact stellar objects. If the acceleration is
sufficiently small, these rings extend quite far from the source and
are reminiscent of what would have been a Keplerian disk in the
absence of acceleration. For larger accelerations, the rings occur
closer to the surface of the star and are more energetic.  This last
property is related to the effect of the extremal acceleration being
at values of $\om$ less than zero.  To see why it is
necessary to study the energy equations first.

The time-like covariant component of the four-velocity, $u_t$,
describes the rest, kinetic, and gravitational energies per unit mass.
Assuming that the acceleration field, needed to hold non-geodesic
circular orbits, is due to a vector potential $\Phi_\rho$, the
corresponding potential energy is calculated by solving Hamilton's
equations~\cite{Jac75,Gol80}:
\begin{equation}
  \label{eq:eqn12}
\frac{dx^\rho}{d\tau} =
\frac{\partial{\cal H}}{\partial \pi_\rho}
\end{equation}
\begin{equation}
  \label{eq:eqn13}
\frac{d\pi_\rho}{
    d\tau} = -\frac{\partial{\cal H}}{\partial
    x^\rho}, 
\end{equation}
where ${\cal H}$ is the super-Hamiltonian, $x^\al$ is a general
coordinate (not be confused with the radial parameter $x$ introduced
in~(\ref{eq:eqn6})) and $\pi_\rho$ is the momentum conjugate to
$x^\rho$.  The super-Hamiltonian for a minimally coupled vector
potential $\Phi_\rho$, is:
\begin{equation}
  \label{eq:eqn14}
{\cal H}={1\over 2m}(\pi_\rho-\Phi_\rho)(\pi^\rho-\Phi^\rho).
\end{equation}
Let the potential be described by a stationary, spherically symmetric scalar 
field in a Kerr space-time, with:
\begin{equation}
  \label{eq:eqn15}
\Phi_\rho=-V(r)\delta^t_\rho+\alpha V(r)\delta^\phi_\rho 
\end{equation}
where $V(r)$ is a real, differentiable, scalar function, depending on
the radial coordinate only since we require stationarity and axial
symmetry, and we confine our attention to the equatorial plane.  In
this case, Hamilton's equations~(\ref{eq:eqn12}) and~(\ref{eq:eqn13}),
lead to:
\begin{eqnarray}
mu_t &= &\pi_t+V   \label{eq:eqn16} \\
mu_k &= & \pi_k+\Phi_k ; \qquad (k=r,\,\theta,\,\phi) \label{eq:eqn17} \\ 
\pi_t &=&{\rm constant}\equiv -E  \label{eq:eqn18} \\
{\dot u_r} &=&-u^tV_{,r}(1-a\Omega) \label{eq:eqn19} \\
\pi_\theta &=&{\rm constant}=0\quad 
{\rm for\,\, equatorial \,\, motion}  \label{eq:eqn20} \\         
\pi_\phi &= & {\rm constant}\equiv m\lambda. \label{eq:eqn21} 
\end{eqnarray}
Here $m$ is the rest mass of the particle, $E$ and $\lambda$ are its total
energy and specific azimuthal angular momentum respectively.  If the
strength $a_{cc}$ of the acceleration field is known, then the
corresponding scaled angular velocity $\om$ can be determined from
equation~(\ref{eq:eqn8}). Written in terms of the parameters
$x,\,\alpha,\,\om$ as in equation~(\ref{eq:eqn6}), equation~(\ref{eq:eqn19})
becomes:
\begin{equation}
  \label{eq:eqn22}
x^3 V(x)_{,x}=m M a_{cc}(\om_{c_+} - \om)^{1/2}
(\om-\om_{c_-})^{1/2} \Lambda^{-1/2}.
\end{equation}
For any fixed value of $\om$, the potential energy $V$ is calculated
by integrating equation~(\ref{eq:eqn22}) provided that the
acceleration field  and a boundary condition are known.
The exact form of the acceleration field will depend on interaction of
the photon field with the  surrounding matter and of that matter with
itself. It is not necessary for us to go into such complexity, for as we
shall see, the exact form of the acceleration field is not
crucial to the results. We shall side-step the whole issue of
determining the acceleration field and simply choose it to be the
following:
\begin{equation}
  \label{eq:eqn22a}
a_{cc} = \frac{\ep x^2\sqrt{\Lam}}{M}.
\end{equation}
The reasons for choosing this form of acceleration field are that, it
is simple, the Newtonian limit is an inverse square law, and  it
simplifies the integration of equation~(\ref{eq:eqn22}).  The
$\sqrt{\Lam}$ term in equation~(\ref{eq:eqn22a}) can be thought of as
due to gravitational red-shift.  

\begin{figure}[htbp]
  \begin{center}
    \includegraphics[width=\textwidth]{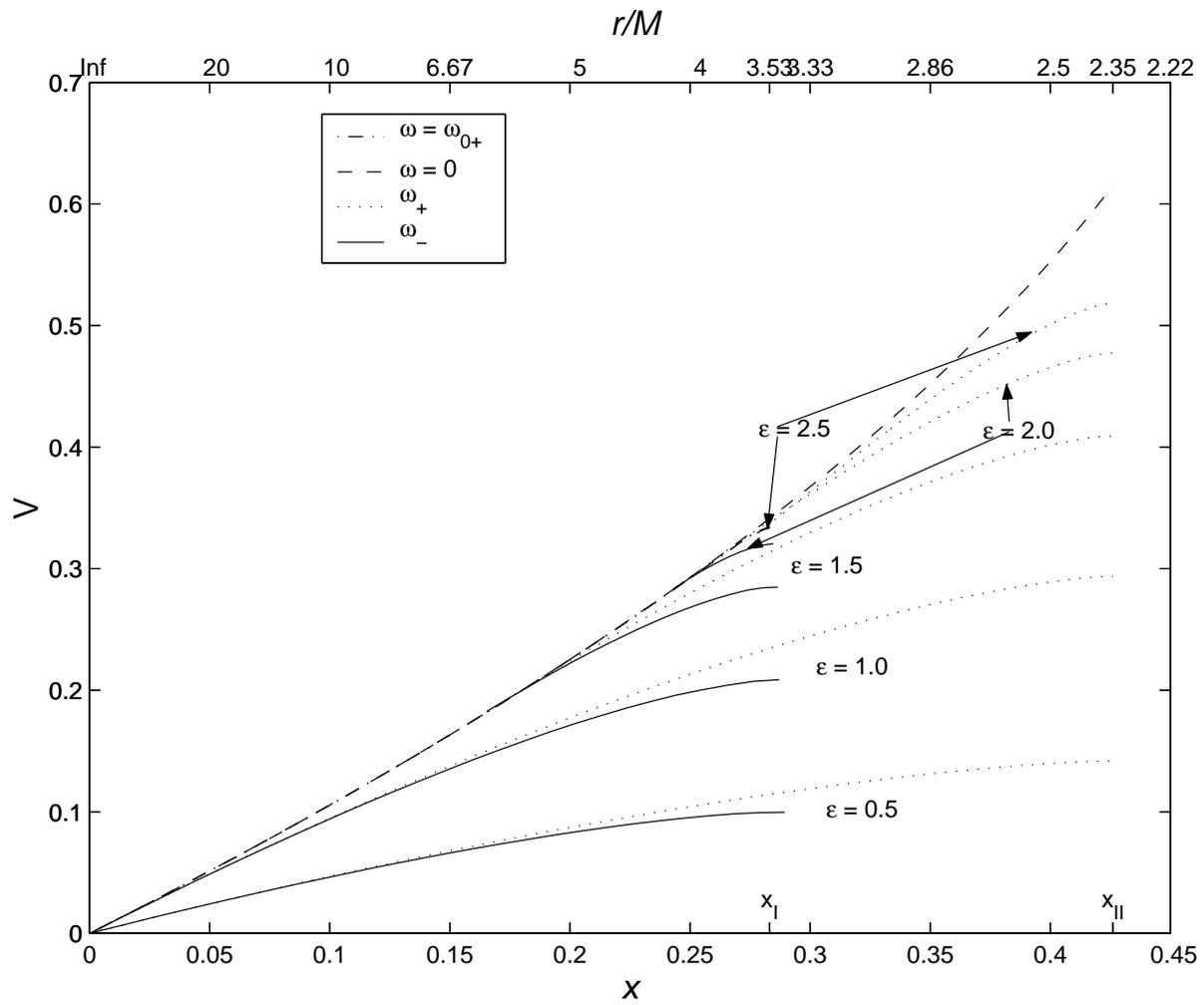}
    \caption[pot.eps]{The potential energy as a function of distance
      for the acceleration field described by~(\ref{eq:eqn22}) for
      $\ep = 0.5, 1.0, 1.5, 2.0,2.5$. \label{fig:pot}}
  \end{center}
\end{figure}

\begin{figure}[htbp]
  \begin{center}
    \includegraphics[width=\textwidth]{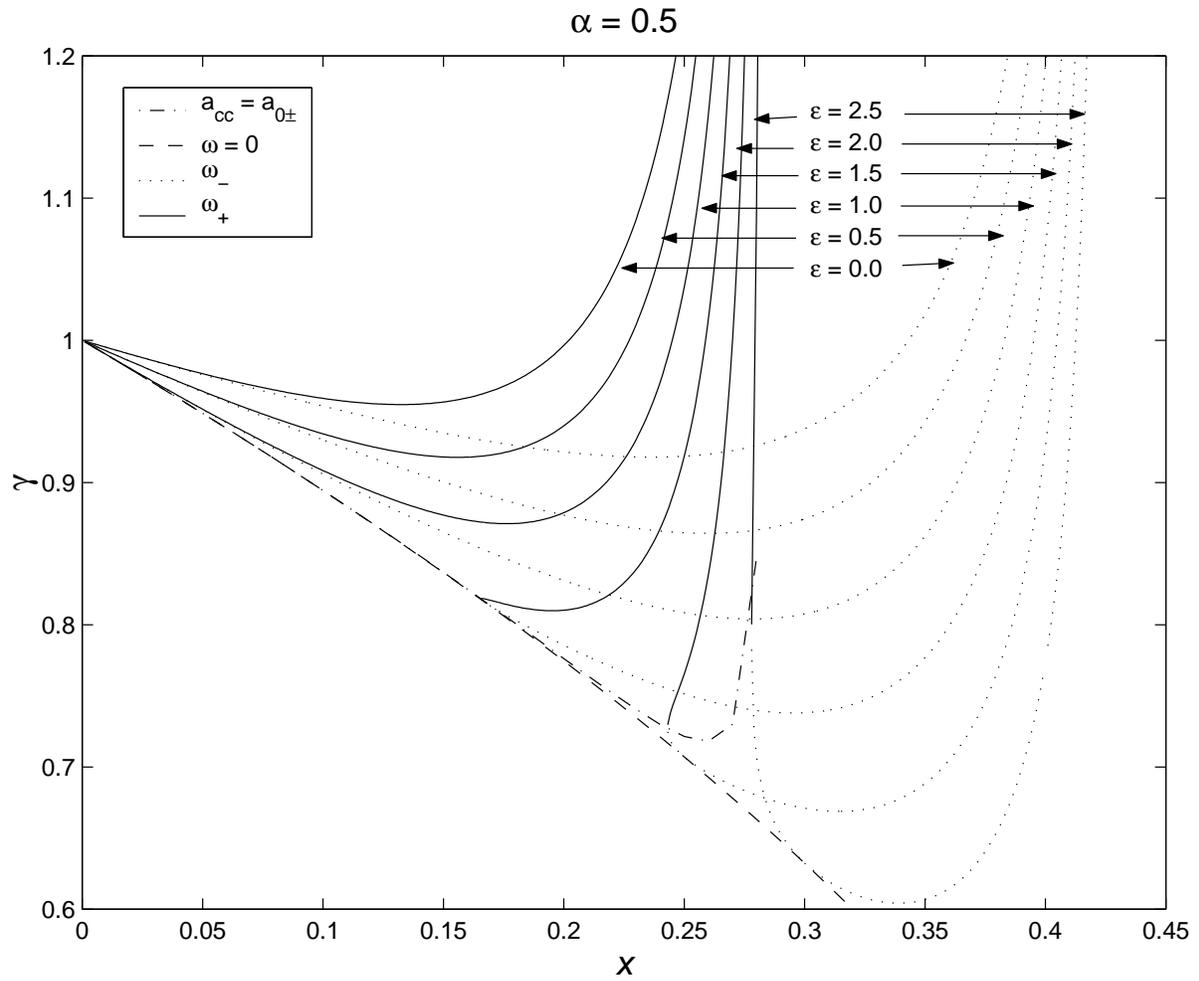}
    \caption{Specific energy as a function
    of $x \equiv M/r$ in an acceleration field given by
    equation~(\ref{eq:eqn22a}) for $\al = 0.5$. 
    \label{fig:gamma}}
  \end{center}
\end{figure}

Figure~\ref{fig:pot} shows the numerical integration of
equation~(\ref{eq:eqn22}) for a unit mass in an acceleration field
described by equation~(\ref{eq:eqn22a}) with the boundary condition
$V(0) = 0$. From this figure we see that the potential energy $V$
levels off at $x_I$ for $\om_-$ orbits and $x_{II}$ for $\om_+$
orbits. 
The total energy $E$ comprises of both, the potential energy $V$ and
the specific energy $mu_t$.  The specific energy is calculated
from~(\ref{eq:eqn1}) and~(\ref{eq:eqn2}),
\begin{equation}
-u_t  =  {x + \al x \om - 2x^2\over \sqrt{x^2 - 2x^3 + 
2\al x^2 \om - \om^2}}\equiv \gamma. \label{eq:eqn23}
\end{equation}
Figure~\ref{fig:gamma} shows the specific energy for the acceleration
field given in equation~(\ref{eq:eqn22a}).

It has been known for quite some time that both stable
counter-rotating and co-rotating geodesics have a minimum
energy~\cite{Fel68}. 
The location of this minimum marks the corresponding last stable
orbit, as circular orbits below this limit require a larger energy.

\begin{figure}[htbp]
  \begin{center}
    \includegraphics[width=\textwidth]{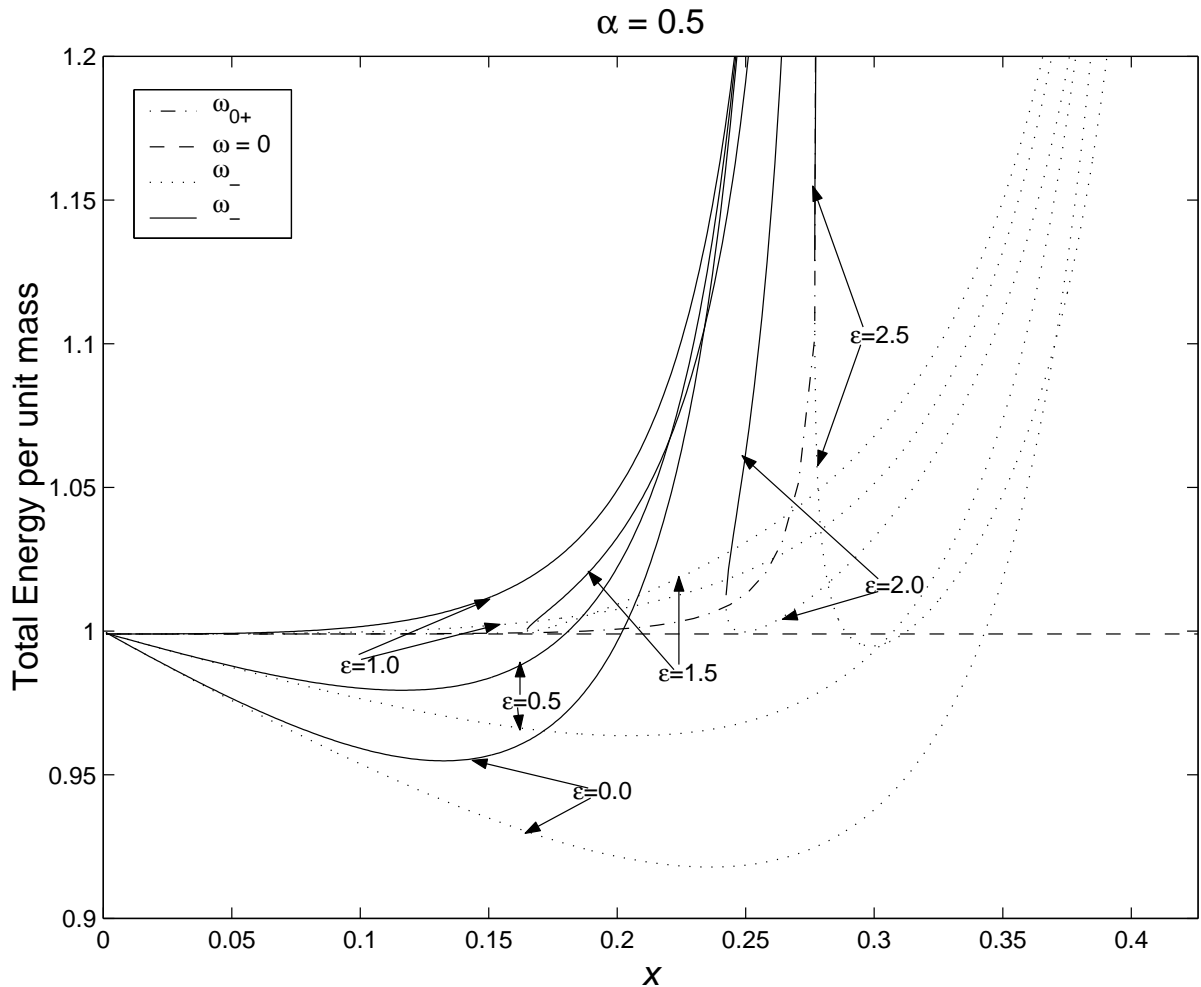}
    \caption{Total energy  as a function distance $x \equiv M/r$
      of circular orbits in the equatorial plane of  a Kerr space-time
      with $\al = 0.5$ and an acceleration field given by
      equation~(\ref{eq:eqn22a}). Plots are for $\ep = 0.0, 0.5, 1.0,
      1.5, 2.0$, and 2.5.  \label{fig:total_eng}}
  \end{center}
\end{figure}

The location of the minimum is determined by solving $\frac{d\ga}{dx}
= 0$ for $\om = \pm x^{3/2}$.  In the presence of an acceleration
field a minimum in energy can still occur, but it is located at
$\frac{dE}{dx} = 0$, where $E$ is the sum of the specific and the
potential energies, $E = m\ga + V$. The location and value of the
minimum of the energy can be determined using
equations~(\ref{eq:eqn22})--(\ref{eq:eqn23}), though in general it is
a rather messy operation. Since the orbits are not geodesics, the
concept of stability here should be handled with care; we mean that a
small loss of the total energy allows the particle to move on a nearby
circular orbit with the same acceleration field. The behaviour of the
particles with respect to a small perturbation of the acceleration
field itself, is matter of a detailed investigation.
Figure~\ref{fig:total_eng} shows energy per unit mass as a function of
$x$ for different acceleration fields. We see from these graphs that
for sufficiently large acceleration the $\om_+$ orbits have a minimum
energy significantly smaller than the energy of the outer orbit. The
reason for this is that while the potential energy increases with $x$
the specific energy of the $\om_+$ orbits drop rapidly before
increasing again. The sudden drop in the specific energy is due to the
sensitivity of the kinetic energy to changes in the modulus of $\om$.
Figure~\ref{fig:wvx} shows that $\om_+$ changes from
negative to zero in a comparatively small region before becoming
positive.   For large accelerations
$\om_+$ at the outer boundary is close to the causal limit hence the
kinetic energy is large. This kinetic energy drops to a minimum as $\om_+$
approaches zero and then increases as $\om_+$ increases. The potential
energy on the other hand does not change significantly and hence a
minimum energy occurs when $\om_+\approx 0$.

In the presence of an outward pointing acceleration field, there
exists, in general, both an inner and an outer boundary. The inner
boundary is determined by the minimum in the energy. An outer boundary
occurs if, at some point, the acceleration field is greater than the
maximal acceleration, since circular orbits exist in region 1 if and
only if $a_{cc} \le a_{0_+}$.  The Mac Lauren expansion of
equations~(\ref{eq:eqn11}) and~(\ref{eq:eqn22a}) both vanish at
infinity, namely at $x=0$. However, for $\ep > 1$ the maximal
acceleration $a_{0+}$ vanishes more quickly, hence, for any given
value of $\epsilon>1$, there are no circular orbits sufficiently far
from the source. As $x$ increases, so does $a_{0+}$.  The point where
$a_{cc}= a_{0+}$, is the outer radius, $x_o$ say, as it is the
smallest value of $x$ (the largest value of $r$) where a circular
orbit is allowed.

If both inner and outer boundaries exist then a ring structure is
formed. 

The efficiency of energy emitted is calculated by comparing the energy
of the outer orbit with the inner one, i.e., 
$${\rm energy\: efficeincy\:}= \frac{E(x_o) - E(x_i)}{E(x_o)}$$
For $\ep =2.5$ the energy efficiency is $11\%$. As $\ep$ increases so
does the efficiency until $a_{cc}(x_I) = a_{0+}(x_I)$, at which point
$E(x_o) = E(x_I) = 0$. 

\section{Astrophysical Implications} \label{sec:ai}
The ring structures carry a large amount of energy which can be converted 
into heat and radiation if accretion takes place. The amount of energy 
which can be released,  
is given by the difference between the total energies of the outer and the 
inner boundaries. 

The counter-rotating rings carry an amount of energy which increases with
$\epsilon$ as shown in figure~\ref{fig:total_eng}.
It is possible  to compare the energy output, in the
case of accretion, from any given non-Keplerian ring of matter, with
that of a Keplerian disk. For the latter case
the outer boundary, with $E_{out}=1$, is at infinity,
while the inner boundary is at $x\approx 0.283$ with $E_{in}\approx
0.917$, for $\al = 0.5$, implying only an $8\%$ efficiency. Although in our case it is
not appropriate to talk about efficiency, since the energy needed to
first generate the acceleration field and set up the ring pattern is
not known, we may just consider the efficiency of the energy release
of a given ring by accretion, say, only once it was formed.

As it is clearly shown by figure~\ref{fig:total_eng}, the inner rings
can carry a large amount of energy which, once released, can
contribute significantly to the total energy output of the
source. Figure~\ref{fig:zerosof10} implies that as 
the star's rotation increases, these energetic rings extend outward
becoming  potentially easier to  observe.  Indeed the effect we
are discussing is sensitive to the rotation parameter $\al$.
 
The space outside a radiating source is  filled with photons. In this case, 
the stress energy tensor of  dust (non interacting matter) in 
the presence of a photon field is 
\begin{equation}
T_{\al\beta} = \rho_d u_\al u_\beta + P_p g_{\al\beta},\label{eq:eqn23a}  
\end{equation}
where $u^\al$ is the four velocity of the dust, $\rho_d$ is its
density and $P_p$ is the photon pressure.  By contracting the
conservation equation ($T^{\beta}_\ga;_\beta = 0$) with the projection
tensor $h_\al^\ga = \de_\al^\ga + u_\al u^\ga$, the acceleration can
be expressed in terms of the pressure gradient~\cite{Ste82}.
If the four velocity is given by equation~(\ref{eq:eqn2}), then
$$
\dot u_\al = -{\partial_\alpha P_{p}\over\rho_d}.
$$
From metric~(\ref{eq:eqn1}) and assuming that the radiation pressure is only radial, then
\begin{equation}
a_{cc}={\Lambda^{1/2}|\partial_r P_p|\over \rho_d}.\label{eq:eqn24}
\end{equation}

\section{Conclusions}\label{sec:c}

In this paper, we have shown that highly energetic
rings of matter can occur around the exterior of compact stars. We
have shown that these rings can achieve energy efficiencies much
greater than those of Keplerian disks. We have also shown that the
size and energy efficiency of these rings depends on, the specific
angular momentum of the star, its mass,  the strength of the
acceleration field it produces, and the properties of the matter with
the ring itself.  

This paper forms the corner stone of work to come. Having established
the general theory behind the formation of non-Keplerian rings there
is still a lot of theoretical and modelling work to be done. For
example, the results of this work make it possible to determine the
energy efficiency as a function of  angular velocity for
a given acceleration field. Once this has been done it is then
possible to deduce the spin rate of a star from the amount of
energy produced by its rings. 

The acceleration field we have studied in this paper is only one of
many possibilities. While the general property of the formation of
rings is not likely to change for different models of the acceleration
the specific nature of the rings, such as their size and efficiency
will. 

The structure of matter within the rings will determine the acceleration
field there. It would be interesting to examine the fields produced by
matter with a polytropic equation of state. 

Similarly, the Hamiltonian describing the particle dynamics was chosen
to be a minimally coupled one, but there are many other options. Even
the choice of vector potential is not unique and more theoretical
work is needed to understand the form it should take. 

The possibilities for extensions of this work, while not endless are
certainly large. The consequences of such work are that it will make it
possible to determine some of the fundamental properties of the stars
by observing the rings around them.

\subsection*{Acknowledgments}
This work was supported by the Ministero
    degli Affari Esteri e dal Ministero della Ricerca Scientifica e
    Tecnologica of Italy.
I wish to thank the Director of the Department of
Physics {\it G. Galilei} of the University of Padova for his
    hospitality during my stay. I would also like to thank Fernando de
    Felice for many helpful discussions and good advice.

\end{document}